\documentclass[lettersize,journal]{IEEEtran}
\usepackage{amsmath,amsfonts}
\usepackage{algorithmic}
\usepackage{array}
\usepackage[caption=false,font=normalsize,labelfont=sf,textfont=sf]{subfig}
\usepackage{textcomp}
\usepackage{stfloats}
\usepackage{url}
\usepackage{verbatim}
\usepackage{graphicx}

\usepackage{booktabs}       
\usepackage{multirow}

\def\etal{\emph{et al.}}

\def\BibTeX{{\rm B\kern-.05em{\sc i\kern-.025em b}\kern-.08em
    T\kern-.1667em\lower.7ex\hbox{E}\kern-.125emX}}
\usepackage{balance}
\begin{document}
\title{TCAN: Text-oriented Cross Attention Network for Multimodal Sentiment Analysis}

\author{Weize Quan, 
        Yunfei Feng,
        Ming Zhou,
        Yunzhen Zhao,
        Tong Wang,
        and Dong-Ming Yan
\thanks{W. Quan and D.-M. Yan are with the State Key Laboratory of Multimodal Artificial Intelligence Systems (MAIS), Institute of Automation, Chinese Academy of Sciences, Beijing 100190, China, and the School of Artificial Intelligence, University of Chinese Academy of Sciences, Beijing 100049, China.}
\thanks{Y. Feng is with the School of Artificial Intelligence and Automation, Huazhong University of Science and Technology, Wuhan 430074, China.}
\thanks{Y. Zhao is with Tencent, Shenzhen 518057, China.}
\thanks{M. Zhou and T. Wang are with the School of Information Science and Technology, Donghua University, Shanghai 200051, China.}
}

\markboth{Journal of \LaTeX\ Class Files,~Vol.~18, No.~9, September~2020}%
{How to Use the IEEEtran \LaTeX \ Templates}

\maketitle

\begin{abstract}
Multimodal Sentiment Analysis (MSA) endeavors to understand human sentiment by leveraging language, visual, and acoustic modalities. Despite the remarkable performance exhibited by previous MSA approaches, the presence of inherent multimodal heterogeneities poses a challenge, with the contribution of different modalities varying considerably. Past research predominantly focused on improving representation learning techniques and feature fusion strategies. However, many of these efforts overlooked the variation in semantic richness among different modalities, treating each modality uniformly. This approach may lead to underestimating the significance of strong modalities while overemphasizing the importance of weak ones. Motivated by these insights, we introduce a Text-oriented Cross-Attention Network (TCAN), emphasizing the predominant role of the text modality in MSA. Specifically, for each multimodal sample, by taking unaligned sequences of the three modalities as inputs, we initially allocate the extracted unimodal features into a visual-text and an acoustic-text pair. Subsequently, we implement self-attention on the text modality and apply text-queried cross-attention to the visual and acoustic modalities. To mitigate the influence of noise signals and redundant features, we incorporate a gated control mechanism into the framework. Additionally, we introduce unimodal joint learning to gain a deeper understanding of homogeneous emotional tendencies across diverse modalities through backpropagation. Experimental results demonstrate that TCAN consistently outperforms state-of-the-art MSA methods on two datasets (CMU-MOSI and CMU-MOSEI).
\end{abstract}


\section{Introduction}
\label{sec:intro}

\begin{figure}[htp]
    \centering
    \includegraphics[width=\linewidth]{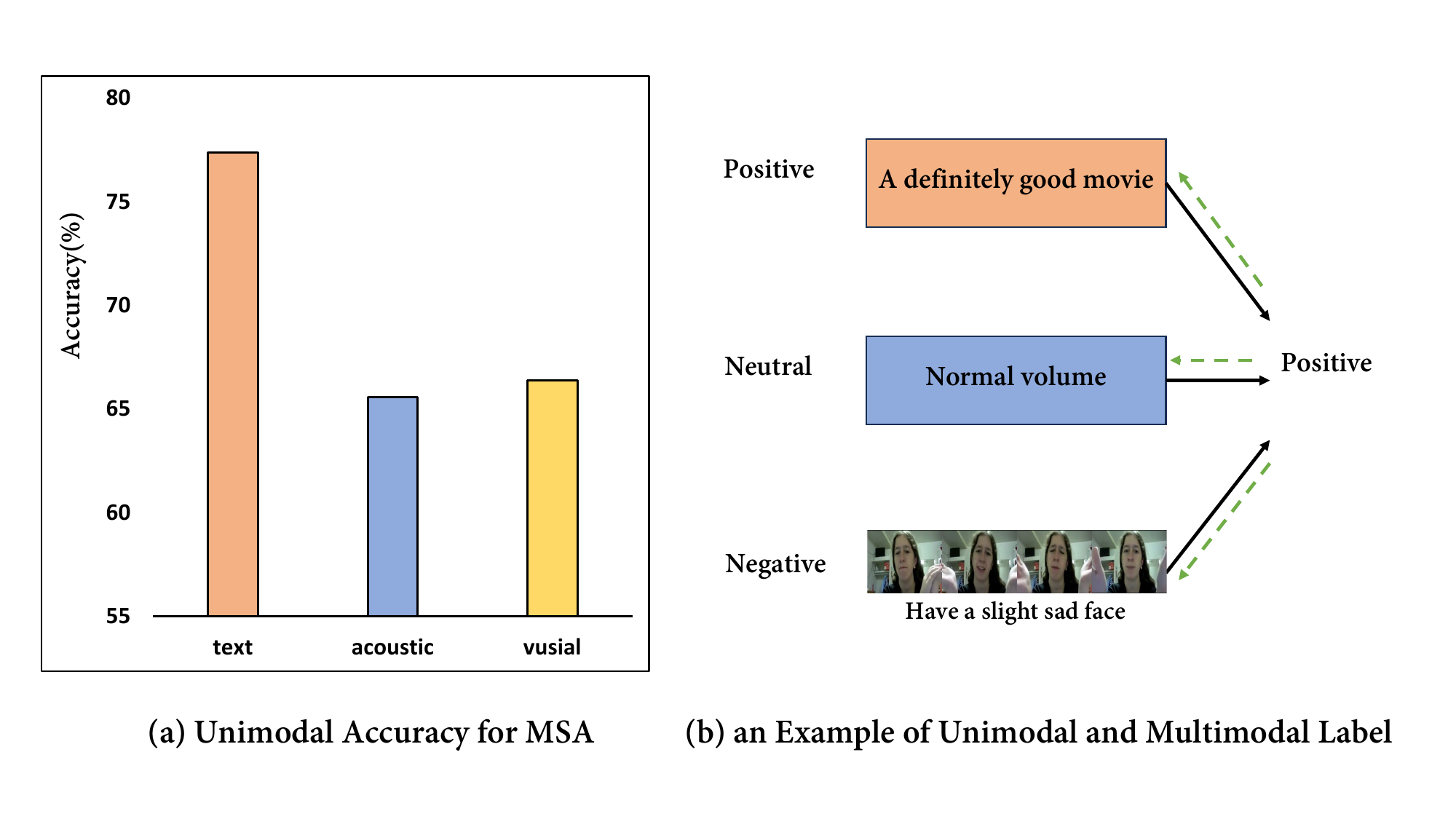}
    \caption{(a) illustrates the significant sentiment analysis discrepancies using unimodality, adapted from Mult \cite{tsai2019multimodal}. (b) shows an example of unimodal labels and multimodal labels, where the green dotted lines represent the process of backpropagation. }
    \label{fig:figure1}
\end{figure}

\IEEEPARstart{I}{n} recent years, the widespread use of social media and the prevalence of smartphones equipped with high-quality cameras have led to a significant increase in multimodal data sharing. This surge includes various forms, such as video clips on diverse social media platforms. Such multimodal data consists of three channels: visual (image), acoustic (voice), and transcribed linguistic (text) data. Various modalities often synergize within the same data segment, offering additional cues that aid in disambiguating semantics and emotions. Recognizing the research significance of sentiment, there has been a notable surge in recent studies \cite{du2021learning,yang2023context,yang2022emotion} focusing on the detection and examination of human sentiments. A majority of these studies leverage Multimodal Sentiment Analysis (MSA), aiming to extract sentiment information from multiple data sources~\cite{zhu2023multimodal,yuan2024noise}. This approach seeks a more comprehensive and precise understanding of a broad spectrum of intricate human emotions, standing in contrast to traditional unimodal sentiment analysis, which typically relies on a single source of information.

Although data from different modalities can be complementary, the redundancy and difficulty in alignment between different modal sequences often pose challenges to fusion. To tackle this issue, previous methods \cite{tsai2018learning,wang2019words} mostly align visual and acoustic sequences from the perspective of text word resolution, but this leads to increased labor costs and ignores long-term dependencies between different modal elements. After the attention mechanism was proposed, many methods like \cite{lv2021progressive,tsai2019multimodal} for dealing with unaligned multimodal sequences are now based on cross-modal attention. These approaches typically involve capturing and utilizing inter-modality correlations through sufficient interactions between each pair of modalities. Nevertheless, this approach causes an increase in the number of parameters and the presence of redundant information across different modalities. Moreover, they assign equal weights to all modalities without considering the varying semantic richness of each modality, potentially neglecting strong modalities and overvaluing weaker ones. Based on our observations of previous works \cite{tsai2019multimodal,chen2017multimodal,pham2019improving}, we notice a dominance of the text modality in MSA. The most significant reason is that the text modality inherently possesses a high level of structure and semantic density. Therefore, it becomes crucial to develop a novel MSA framework that places importance on the text modality while striking a balance in the contributions made by different modalities.

Taking inspiration from the aforementioned observations, we propose a Text-oriented Cross-Attention Network (TCAN) as an effective means to refine multimodal representations. The fusion scheme involves two bi-modal fusion modules, which is distinctly different from the traditional ternary symmetric approach. As illustrated in Fig.~\ref{fig:figure1}(a), based on empirical evidence highlighting the significance of the text modality, our model incorporates two text-related modality pairs, TV (text-visual) and TA (text-acoustic), as inputs for its respective bi-modal learning modules. It then encourages modalities to iteratively enhance their information by engaging in interactive learning with their respective counterparts. In order to maintain fairness of the bidirectional learning process in both modalities, the two learning networks in our proposed model are exactly the same. Our model is built upon a fundamental framework consisting of stacked transformers, which have demonstrated their effectiveness in the realm of multimodal learning \cite{yu2020improving}.

Nonetheless, the vanilla Transformer \cite{vaswani2017attention} does possess certain limitations. The cross-attention mechanism introduces redundancy and noise while directing attention to information within the modality, particularly for the visual and acoustic modalities \cite{lei2024text,han2021bi}. Since acoustic and visual modalities usually require pre-processing \cite{chen2017multimodal} before being input to the network, and there will be a certain bias during the pre-processing process, the impact of noise caused by bias is often unavoidable. Furthermore, the redundancy in time series between visual and acoustic sequences is remarkably high. Simple fusion and splicing of the fused features would introduce interference from noise and redundant information. To address these challenges, we propose a gated-control module that incorporates two gates. Different from TMRN \cite{lei2024text}, which uses the same gate, we train two separate gates in the network, and each gate is responsible for different tasks. The first gate determines the proportion of the target modality's components to be forwarded, while the second gate determines the proportion of compound components to be injected into the target modality. This design aims to enhance the extraction of semantic information from time series, effectively filtering out noise in visual and acoustic modalities.

In this paper, we simultaneously exploit forward-guidance and backward-guidance approaches. Inspired by methods utilizing independent annotations for each modality and the advancements in modality-specific representation learning, we introduce unimodal joint training during the training process. Unlike the previous method \cite{yu2021learning}, we avoid generating unimodal labels and instead directly predict multimodal labels. As shown in Fig.~\ref{fig:figure1}(b), unimodal labels and multimodal labels may be different or even opposite. To address this disparity, we propose a shared-weight encoder to jointly train single-modal label prediction for all three modalities, encouraging the network to learn cross-modal homogeneousness in a backward-guiding manner. 

The contributions of this work can be summarized as:

\begin{itemize}
    \item We propose a novel network called TCAN, focusing on the dominance of the text modality in MSA tasks. TCAN leverages cross-attention and the gate network to reinforce the interaction between text modality and the other two modalities, with the text modality serving as the main thread to attain denoised feature representations with low redundancy.
    \item We introduce a shared-weight encoder to extract cross-modal homogeneous features in multimodal joint training, obtaining better multimodal fusion features through backpropagation. 
    \item Comprehensive experiments show that our method outperforms previous state-of-the-art methods on two widely used MSA benchmarks (CMU-MOSI and CMU-MOSEI).
\end{itemize}

\section{Related Work}

\label{sec:related}
In this section, we review several closely related topics, including multimodal sentiment analysis, multimodal representation learning, and joint learning.

\subsection{Multimodal Sentiment Analysis}
Multimodal Sentiment Analysis (MSA) is to deduce human sentiment by analyzing the language, visual, and acoustic elements present in video clips. MSA can benefit from the diverse nature of information provided by different modalities, which can offer varying levels of insight into human sentiment. There are two main categories of approaches in Multimodal Sentiment Analysis: fusion strategy-based methods \cite{liu2018efficient,zadeh2017tensor,zadeh2018memory,huang2024dominant} and cross-modal attention-based methods \cite{liang2021attention,lv2021progressive,tsai2019multimodal,wang2023cross}.

The former focuses on developing advanced strategies for multimodal fusion to create discriminative multimodal representations, Zadeh \etal~\cite{zadeh2017tensor} introduced a Tensor Fusion Network (TFN) that can aggregate unimodal, bimodal, and trimodal interactions progressively. To solve the problem of excessive parameters in the fusion model, Liu \etal~\cite{liu2018efficient} proposed the Low-rank Multimodal Fusion (LMF) method, which uses weights to perform low-rank matrix decomposition. However, the fusion of multimodal features is often challenged due to the inherent heterogeneity and redundant information present across different modalities. To address this challenge, some research \cite{hazarika2020misa, tsai2018learning, yang2022disentangled} focuses on exploring the characteristics and commonalities of multimodal representations by decoupling the features. This approach aims to facilitate a more effective fusion of multimodal representations. Hazarika \etal~\cite{hazarika2020misa} proposed a method that decomposes the multimodal features into modality-invariant and modality-specific components. It enables the learning of refined multimodal representations. DMD \cite{li2023decoupled} decoupled the multimodal features into a homogeneous (modality-irrelevant) part and a heterogeneous (modality-exclusive) part. By decoupling the features, the redundant information is reduced, and a comprehensive perspective of the multimodal data is obtained.

In recent times, the development of MSA has been driven by cross-modal attention-based approaches. These methods focus on learning cross-modal correlations to enhance the representation of each modality involved. MulT \cite{tsai2019multimodal} is an example of a cross-modal attention-based approach that has become a representative work in the field of Multimodal Sentiment Analysis. This approach used a transformer-based architecture and employed various attention mechanisms to model the interactions between different modalities, leading to improved sentiment analysis performance. Han \etal~\cite{han2021bi} changed the ternary symmetric approach into two bi-modal fusion modules. Lv \etal~\cite{lv2021progressive} introduced a progressive modality reinforcement method based on \cite{tsai2019multimodal}, it aims to extract and understand the inherent connections and correlations between multimodal representations and their corresponding unimodal representations. Similar to the aforementioned approaches \cite{han2021bi, tsai2019multimodal}, our work also leverages the attention mechanism as a fundamental component. However, in our work a gating mechanism is introduced on this basis to filter out the influence of redundant features and noise.

\subsection{Multimodal Representation Learning}
\label{sec:2.2}
The goal of multimodal representation learning is to acquire cohesive representations that capture the shared information across various modalities, including but not limited to images, text, and audio. The objective is to establish a unified representation space that facilitates the integration of diverse modalities by establishing meaningful connections based on their semantic similarities. Multimodal representation learning can be achieved through two primary approaches, outlined as follows: 

One of the two primary approaches is a correlation-based approach. Correlation is considered a key measure of commonalities between multiple modalities. Numerous prior studies have incorporated correlation analysis in their research and achieved good results. Sun \etal~\cite{sun2019multi} introduced a two-step Deep Canonical Correlation Analysis (DCCA) framework to learn multimodal representation. ICCN \cite{sun2020learning} also learned correlations between modalities via DCCA, which is an example of correlation-based methods. Mittal \etal~\cite{mittal2020m3er} introduced a selection criterion that uses Canonical Correlational Analysis to differentiate between ineffective and effective modalities. Although considering the correlation between all modalities, these works neglected the different impact of each modality on the final result, that is, the competition between modalities.

The other approach is the alignment-based approach. Alignment refers to the process of mapping signals with different sampling rates to a common frequency. Traditionally, early methods \cite{blikstein2016multimodal,zadeh2016multimodal} for multimodal alignment typically involved selecting a specific target frequency and then determining the frames in each modality that required mapping to that particular position. Loss functions such as CTC \cite{graves2006connectionist}, InfoNCE \cite{he2020momentum} and its variations are commonly employed to enhance the improvement of alignment in various applications. After the attention mechanism was proposed, the transformer framework achieved extraordinary results in many fields, including visual and text fields \cite{dosovitskiy2020image,vaswani2017attention}. Different from traditional alignment methods based on spatial distance, attention mechanisms enable a more flexible and effective soft alignment between two modalities. This approach has been demonstrated to be effective in a broader range of multimodal representation learning and feature fusion scenarios. For example, Tsai \etal~\cite{tsai2019multimodal} designed a transformer-based architecture and employed various attention mechanisms to align different modality sequences, leading to improved sentiment analysis performance. Yu \etal~\cite{yu2020improving} proposed a Unified Multimodal Transformer to address the MSA problem, which was designed to simultaneously model text and visual representations. The fusion framework used in our proposed method is based on the Transformer, and on this basis, we also introduce a gating mechanism to reduce the impact of noise and redundant features.

\subsection{Joint Learning}
Joint learning is also a way of multimodal feature learning, in which we can enrich multimodal features through single-modal feature learning. Different from the methods mentioned in Section \ref{sec:2.2}, these approaches are usually based on backward guidance. However, due to the unified multimodal annotation, it is difficult for them to directly utilize multimodal labels to participate in training. Unimodal joint learning usually has the following two methods: additional unimodal annotations and auto-generated unimodal labels. The former like Yu \etal~\cite{yu2020ch} introduced fine-grained annotation of modality, which needed high labor costs. An example of the latter method is \cite{yu2021learning} which designed a unimodal label generation module based on multimodal labels and modality representations. Different from both the above methods, we propose a homogeneous encoder with shared weights. Through this encoder, we can extract unimodal homogeneous features to participate in unimodal joint training.

\begin{figure*}[ht]
    \centering
    \includegraphics[width=\linewidth]{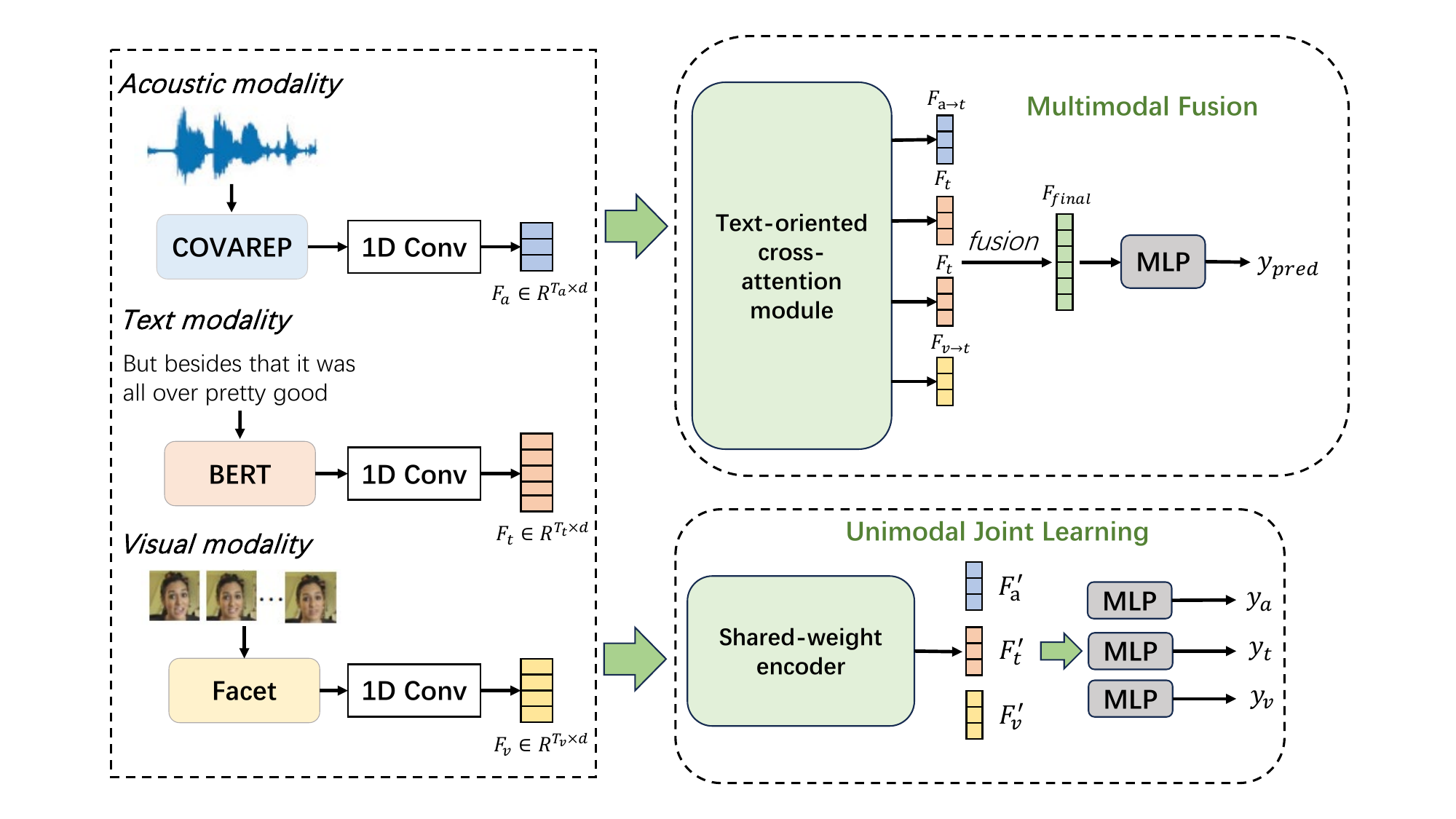}
    \caption{The framework of TCAN. Given the input multimodal data, TCAN encodes their respective shallow features $F_{m}$, where $ m\in \left \{ t,v,a \right \}$. In the Text-oriented Cross-attention module, TCAN exploits cross-attention and self-attention mechanisms to process text-audio pairs and text-video pairs and applies the gated mechanism to control the impact of noise and redundant information (Section \ref{sec:3.3}). At the same time, we introduce a shared-weight encoder called the Homogeneous encoder, which is used to extract homogeneous features from unimodal for joint training (Section \ref{sec:3.4}). Finally, we compute the final representation from the outputs of the last layer in the Text-oriented Cross-attention module and then concatenate them for MSA.}
    \label{fig:framework}
\end{figure*}

\section{Method}
\label{sec:method}

The overall framework of our TCAN is shown in Fig.~\ref{fig:framework}, which contains three components: 1) Unimodal feature extraction module: this module focuses on generating feature representations for input data. We leverage pre-trained BERT \cite{devlin2018bert} for text data, while visual and acoustic features are extracted using pre-trained Facet \cite{baltruvsaitis2016openface} and COVAREP \cite{degottex2014covarep}, respectively. 2) Text-oriented cross-attention network: this section employs modality fusion and gate mechanism to facilitate the interaction among unimodal features. Categorizing the extracted features into visual-text and acoustic-text pairs, we implement self-attention on the text modality and apply text-queried cross-attention to the visual and acoustic modalities. To eliminate noise and redundant representations, a gated mechanism with two gates is incorporated into the attention modules. 3) Unimodal joint learning: in addition to the above multimodal prediction branch, the unimodal data from the three modalities are collectively trained to predict single-modal labels with a shared-weight encoder. This encourages the network to learn cross-modal consistency through backpropagation.

\subsection{Problem Statement}
The multimodal sentiment analysis task in this work uses the text ($t$), visual ($v$), and acoustic ($a$) information from the same video clip as inputs, which are represented as $X_{m}\in  R^{T_{m}\times d_{m} }$ for each modality $ m\in \left \{ t,v, a \right \}$. Throughout the following text, the notations $T_{m}$ and $d_{m}$ are employed to denote sequence length and feature dimension of modality $m$, respectively. Our model aims to thoroughly explore and integrate sentiment-related information from these input unaligned multimodal sequences. The ultimate objective is to generate a text-driven multimodal representation, which can be utilized to predict the sentiment analysis results accurately.

\subsection{Unimodal Feature Extraction Module}
In order to attain enhanced feature representations for the text modality, we employ a pre-trained BERT \cite{devlin2018bert} to extract sentence-level features:
\begin{equation}
    F^{bert}_t = \mathbf{BERT}(X_{t}; \theta^{bert}_t),
\end{equation}
where $\theta_{t}^{bert}$ means the parameters of BERT. In the visual and acoustic modalities, we follow previous methods \cite{yang2022learning,li2023decoupled} to utilize pre-trained models (Facet \cite{baltruvsaitis2016openface} and COVAREP \cite{degottex2014covarep}) for extracting initial features $F^{facet}_v$ and $F^{cova}_a$ from raw input data. Subsequently, we apply a one-dimensional convolution network (1D) to capture the temporal characteristics and project these features into a fixed dimension as $F_{m}\in R^{T_{m}\times d}$, where $m \in \left \{ t,v,a \right \}$.
\begin{equation}
\begin{split}
    F_{a} &= \mathbf{1D}_a(F^{cova}_a; \theta^{1D}_a), \\
    F_{v} &= \mathbf{1D}_v(F^{facet}_v; \theta^{1D}_v), \\
    F_{t} &= \mathbf{1D}_v(F^{bert}_t; \theta^{1D}_t),
\end{split}
\end{equation}
where $\theta_{m}^{1D}$ refers to the parameters of one-dimensional convolutional network.

\subsection{Text-oriented Cross-Attention Module}
\label{sec:3.3}
\begin{figure}[t]
    \centering
    \includegraphics[width=\linewidth]{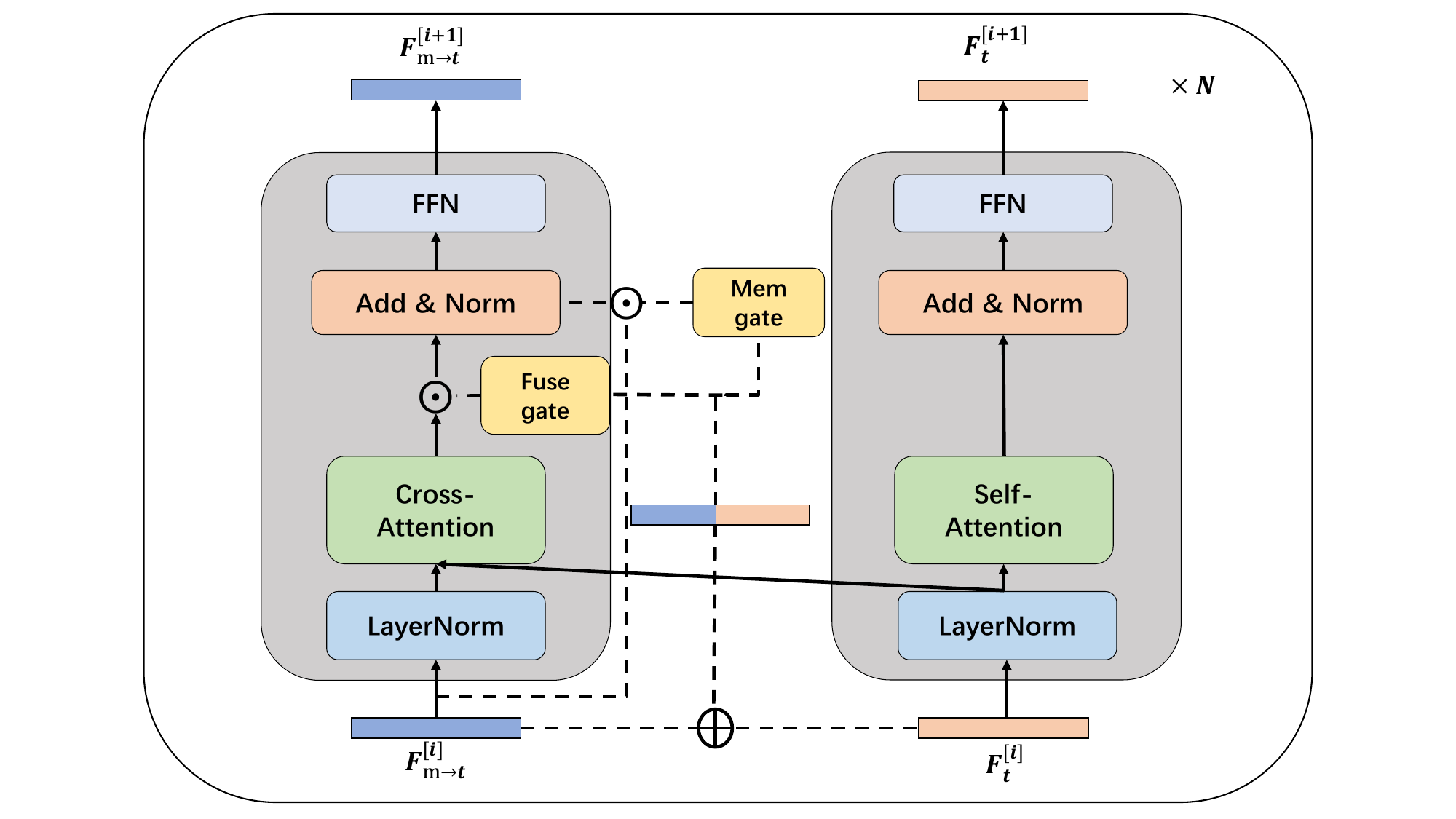}
    \caption{The architecture of the Text-oriented Cross-attention module. The cross-attention block takes text-audio and text-video pairs as its input and the self-attention block takes text as its input. After the cross-attention block, two gates were introduced to eliminate adverse information. The memory gate decides how much proportion of the target modality's components to be kept forwarding, and the fuse gate decides how much proportion of fused components to be injected into the target modality. }
    \label{fig:unit}
\end{figure}

The architecture of this module is shown in Fig.~\ref{fig:unit}. In our TCAN, text-oriented cross-attention modules are stacked $N$ times. We also evaluate the impact of stack number on the final results in the ablation experiment. The module contains a cross-attention block for text-oriented feature fusion and a self-attention block for refining text features. Additionally, a gate mechanism is incorporated to effectively integrate useful information and filter out redundancies between these two blocks. The cross-attention block takes $F_{t}^{[i]}$ and $F_{m \to t}^{[i]}$ as its inputs, where $m \in \left \{ a,v \right \}$, and the superscription $i$ indicates the \emph{i}-th layer. First, we perform a layer normalization (LN) on the features: $F_{m \to t}^{[i]} = \mathbf{LN}(F_{m \to t}^{[i]})$ and $F_{t}^{[i]} = \mathbf{LN}(F_{t}^{[i]})$. Then, we put them into a cross-attention (CA) block:
\begin{equation}
    \begin{aligned}
       F_{m\to t}^{[i+1]} &= \mathbf{CA}_{m\to t}^{[i]}(F_{m\to t}^{[i]},F_{t}^{[i]}) \\
                        &= softmax(\frac{F_{t}^{[i]}W_{Q_{t}}W_{K_{m}}^{T}{F_{m\to t}^{[i]}}^T}{\sqrt{d} } )F_{m\to t}^{[i]}W_{V_{m}},
    \end{aligned}
\end{equation}
where $F_{m\to t}^{[0]} = F_{m}$, $F_{t}^{[0]} = F_{t}$, and $W_{*}$ means a learnable weight matrix through which the corresponding query, key, and value can be obtained. And the self-attention (SA) block takes $F_{t}^{[i]}$ as input to obtain $F_{t}^{[i+1]}$:
\begin{equation}
    \begin{aligned}
        F_{t}^{[i+1]} &= \mathbf{SA}_{t}^{[i]}(F_{t}^{[i]}) \\
                      &= softmax(\frac{F_{t}^{[i]}W_{Q_{t}}W_{K_{t}}^{T}{F_{t}^{[i]}}^T}{\sqrt{d}})F_{t}^{[i]}W_{V_{t}},
    \end{aligned}
\end{equation}
where $W_{*_{t}}$ represents a learnable weight matrix related to query, key, and value of text modality.

Exclusive reliance on the aforementioned blocks for feature fusion may introduce interference from noise and redundancy through simple attention computation. To address this, we introduce two gates to mitigate the influence of unreliable information. We can use these two gates to adaptively determine the proportion of $F_{m\to t}^{[i]}$ and $F_{m\to t}^{[i+1]}$ that can pass through to achieve features that are more beneficial to the final prediction. The memory gate $g_m$ determines the proportion of the target modality's components to be forwarded, while the fuse gate $g_f$ decides the proportion of fused components to be injected into the target modality. These gate signals are generated from the sequential representations of the two modalities within the same layer:
\begin{equation}
\begin{split}
     g_{m}^{[i]} &= \sigma(W_{m}^{[i]} \cdot [F_{t}^{[i]} \oplus F_{m\to t}^{[i]}]), \\
    g_{f}^{[i]} &= \sigma(W_{f}^{[i]} \cdot [F_{t}^{[i]} \oplus F_{m\to t}^{[i]}]),
\end{split}
\end{equation}
where $W_{*}^{[i]} \in R^{2d\times d}$ is the projection matrix and the subscripts $m$ and $f$ are used to distinguish the two gates, $\sigma$ denotes the sigmoid non-linearity function and $\oplus$ represents concatenation. Finally, the output features can be written as:
\begin{equation}
    F_{m\to t}^{[i+1]} = g_{m}^{[i]} \odot F_{m\to t}^{[i]} + g_{f}^{[i]} \odot F_{m\to t}^{[i+1]},
\end{equation}
where $\odot$ means component-wise multiplication.

This operation can filter the noisy information and redundant features produced by the cross-modal interactions while learning to measure the fusion ratio of two modalities. After that, following the feed-forward structures in transformers, we process $F_{m\to t}^{[i+1]}$ and $F_{t}^{[i+1]}$ by a position-wise Feed Forward Network (FFN) with residual connections:
\begin{equation}
\begin{split}
        F_{m\to t}^{[i+1]} &= \mathbf{FFN}(\mathbf{LN}( F_{m\to t}^{[i+1]})) +  F_{m\to t}^{[i+1]}, \\
    F_{t}^{[i+1]} &= \mathbf{FFN}(\mathbf{LN}(F_{t}^{[i+1]})) + F_{t}^{[i+1]}.
\end{split}
\end{equation}

We calculate the final feature representation by concatenating the outputs of the last layer in each module. Each module comprises two heads, resulting in a total of four heads for concatenation. By combining the outputs of the four heads from the two text-oriented cross-modal attention modules, the final representation can be obtained as follows:
\begin{equation}
    F_{final} = F_{a\to t} \oplus F_{t} \oplus F_{v\to t} \oplus F_{t}.
\end{equation}
Subsequently, the fused feature, denoted as $F_{final}$ is fed into a Multi-Layer Perceptron (MLP) to predict the final sentiment label $y_{pred}$:
\begin{equation}
    y_{pred} = \mathbf{MLP}(F_{final}; \theta^{mlp}_{final}), 
\end{equation}
where $\theta^{mlp}_{final}$ means the parameters of MLP.

\subsection{Unimodal Joint Learning}
\label{sec:3.4}
Each modality has a different impact on the prediction results, and the sentiment of some modalities is opposite to the ground truths. For example, a multimodal sample with a positive label can be negative in the visual modality. To gain a deeper insight into the homogeneous emotional tendencies across diverse modalities, we introduce a unimodal joint learning module in addition to the above multimodal branch. This approach helps encourage the cross-attention network to understand the latent sentiment semantic of each modality more profoundly through backpropagation. Specifically, taking the extracted unimodal features as inputs, we employ a shared-weight encoder $\epsilon$ to enable single-modal prediction for all three modalities:  
\begin{equation}
    F'_{m} = \epsilon (F_{m};\theta_{\epsilon}),
\end{equation}
where $m \in \left \{ t,v,a \right \}$, and $\theta_{\epsilon}$ refers to the parameters of the share-weight encoder. The encoder is a one-dimensional convolution network in our implementation. Once we obtain the homogeneous features from all unimodal features, they are fed into an MLP to generate unimodal predictions denoted as $y_{m}$:
\begin{equation}
    y_{m} = \mathbf{MLP} (F'_{m}; \theta^{mlp}_m), 
\end{equation}
where $\theta^{mlp}_m$ means the parameter of MLP. Note that this strategy is only performed during the training phase. 
\subsection{Optimization Objective}
The optimization objective is based on L1 loss, and the loss function for our multimodal prediction is formulated as follows:
\begin{equation}
    \mathcal{L}_{multi} = \frac{1}{K} \sum_{k=1}^{K} |y^k_{pred}-y^k_{gt}|,
\end{equation}
where $K$ is the number of multimodal samples. $y^k_{pred}$ and $y^k_{gt}$ represents the predicted and ground-truth label of the \emph{k}-th sample respectively. Similarly, in the unimodal joint learning part, the loss can be written as:
\begin{equation}
    \mathcal{L}_{uni} = \frac{1}{K} \sum_{k=1}^{K} \sum_{m \in \{t,v,a\}} |y^k_{m}-y^k_{gt}|,
\end{equation}
where $y^k_{m}$ is the predicted unimodal label for modality $m$ in the \emph{k}-th sample.

In summary, the total loss can be calculated by:
\begin{equation}
    \mathcal{L}_{total} = \mathcal{L}_{multi} + \lambda \mathcal{L}_{uni}, 
\end{equation}
where $\lambda$ is a tunable hyperparameter.

\section{Experiments}
\label{sec:experiment}
In this section, we illustrate our experimental settings including datasets and implementation details, compare our method with advanced competitors, and conduct ablation studies and visualization to deeply analyze our model.

\subsection{Experimental Settings}

\textbf{Datasets}. We evaluate TCAN on CMU-MOSI \cite{zadeh2016mosi} and CMU-MOSEI \cite{zadeh2018multimodal} datasets. To achieve a more comprehensive comparison, the experiments are conducted under word-unaligned settings. Table~\ref{tab:template} depicts the composition of these two datasets. The CMU-MOSI dataset comprises 2,199 short monologue video clips. The acoustic features are extracted at a sampling rate of 12.5 Hz, while the visual features are extracted at a sampling rate of 15 Hz. 
The CMU-MOSEI dataset consists of a significantly larger number of samples compared to CMU-MOSI, with a total of 22,856 movie review video clips sourced from YouTube (approximately 10 times the size of CMU-MOSI). The acoustic features in CMU-MOSEI are extracted at a sampling rate of 20 Hz, while the visual features are extracted at a sampling rate of 15 Hz. 
For both CMU-MOSI and CMU-MOSEI datasets, each sample is annotated with a sentiment score ranging from -3 to 3. The sentiment scores represent different levels of sentiment, including highly negative, negative, weakly negative, neutral, weakly positive, positive, and highly positive. Following previous works \cite{mittal2020m3er, lv2021progressive}, we evaluate the model performance with the following metrics: the 7-class accuracy (Acc-7), the binary accuracy (Acc-2) where zero excluded evaluation results are used, the mean absolute error (MAE), the correlation of the model’s prediction with human (Corr), and the F1 score.

\textbf{Implementation details}. On the two datasets, we exploit a BERT-base-uncased pre-trained model \cite{devlin2018bert} to obtain a 768-dimensional hidden state as the word features. For visual modality, Facet \cite{baltruvsaitis2016openface} was utilized to encode each video frame and detect the presence of all 35 facial action units \cite{li2019self,li2020learning}. To extract features from the acoustic modality, COVAREP \cite{degottex2014covarep} was employed and processed to generate a set of 74-dimensional features. The hyperparameter $\lambda$ in our experiment is set to 0.5. All experiments were conducted using PyTorch on an RTX 3090 GPU with 24GB memory. The training batch size was set to 16, and the TCAN model was trained for 50 epochs until convergence.

\begin{table}[t]
\centering
\caption{
The compositions of the number of samples in the CMU-MOSI and the CMU-MOSEI dataset.
}
\begin{tabular}{lccc}
\toprule
Datasets & \#Train & \#Validation & \#Test \\
\midrule
CMU-MOSI & 1,284 & 229 & 686 \\
CMU-MOSEI & 16,326 & 1,871 & 4,659 \\
\bottomrule
\end{tabular}

\label{tab:template}
\end{table}

\begin{table}[t]
\centering
\caption{
Comparison results on the CMU-MOSI. \textbf{Bold} is the best.
}
\resizebox{\columnwidth}{!}{

\begin{tabular}{lcccccc}
\toprule
Method & $MAE\downarrow$ & $Corr\uparrow$ & $Acc_{7}\uparrow$ & $Acc_{2}\uparrow$ & $F_{1}\uparrow$\\
\midrule
TFN \cite{zadeh2017tensor} & 0.901 & 0.698 & 34.9 & 80.8 & 80.7 \\
LMF \cite{liu2018efficient} & 0.917 & 0.695 & 33.2 & 82.5 & 82.4 \\
MulT \cite{tsai2019multimodal} & 0.861 & 0.711 & - & 84.1 & 83.9 \\
MISA \cite{hazarika2020misa} & 0.783 & 0.761 & 42.3 & 83.4 & 83.6 \\
MAG-BERT \cite{rahman2020integrating} & 0.731 & 0.789 & - & 84.3 & 84.3 \\
Self-MM \cite{yu2021learning} & 0.718 & 0.796 & 45.8 & 84.45 & 84.44 \\
MFSA \cite{yang2022learning} & 0.856 & 0.722 & 41.4 & 83.3 & 83.7 \\
DMD \cite{li2023decoupled} & 0.752 & - & 46.06 & 83.23 & 83.29 \\
DLF \cite{wang2025dlf} & 0.731 & 0.781 & \textbf{47.08} & 85.06 & 85.04 \\
TCAN (Ours) & \textbf{0.714} & \textbf{0.797} & 46.79 & \textbf{86.28} & \textbf{86.15} \\
\bottomrule
\end{tabular}

}

\label{tab:table2}
\end{table}

\begin{table}[t]
\centering
\caption{
Comparison results on the CMU-MOSEI dataset. \textbf{Bold} is the best.
}
\resizebox{\columnwidth}{!}{

\begin{tabular}{lcccccc}
\toprule
Method & $MAE\downarrow$ & $Corr\uparrow$ & $Acc_{7}\uparrow$ & $Acc_{2}\uparrow$ & $F_{1}\uparrow$\\
\midrule
TFN \cite{zadeh2017tensor} & 0.593 & 0.700 & 50.2 & 82.5 & 82.1 \\
LMF \cite{liu2018efficient} & 0.623 & 0.677 & 48.0 & 82.0 & 82.1 \\
MulT \cite{tsai2019multimodal} & 0.580 & 0.703 & - & 82.5 & 82.9 \\
MISA \cite{hazarika2020misa} & 0.568 & 0.724 & - & 84.23 & 83.97 \\
MAG-BERT \cite{rahman2020integrating} & 0.539 & 0.753 & - & 85.2 & 85.1 \\
Self-MM \cite{yu2021learning} & 0.536 & 0.763 & 52.8 & 84.95 & 84.85 \\
MFSA \cite{yang2022learning} & 0.574 & 0.724 & 53.2 & 83.8 & 83.6 \\
DMD \cite{li2023decoupled} & 0.543 & - & 52.78 & 84.62 & 84.62 \\
DLF \cite{wang2025dlf} & 0.536 & 0.764 & \textbf{53.90} & 85.42 & 85.27 \\
TCAN (Ours) & \textbf{0.532} & \textbf{0.774} & 53.10 & \textbf{86.27} & \textbf{86.17} \\
\bottomrule
\end{tabular}
}

\label{tab:table3}
\end{table}


\subsection{Comparisons with the State-of-the-Art}
The proposed approach is compared to the existing state-of-the-art methods, including TFN \cite{zadeh2017tensor}, LMF \cite{liu2018efficient}, Mult \cite{tsai2019multimodal}, MISA \cite{hazarika2020misa}, MAG-BERT \cite{rahman2020integrating}, Self-MM \cite{yu2021learning}, MFSA \cite{yang2022learning}, DMD \cite{li2023decoupled}, and DLF~\cite{wang2025dlf}. Table~\ref{tab:table2} and Table~\ref{tab:table3} show the corresponding results on the CMU-MOSI and CMU-MOSEI datasets, respectively. Our approach demonstrates remarkable superiority over the majority of previous methods \cite{tsai2019multimodal,hazarika2020misa}, showcasing substantial performance improvements across all metrics on both datasets. Compared with the traditional feature fusion method, we added a gating mechanism during modal fusion to filter out the influence of noise and redundant features, and we also introduced unimodal joint training. Therefore, our method achieved relatively good results. Moreover, our model exhibits superior or competitive performance compared to the current state-of-the-art DLF~\cite{wang2025dlf} across most metrics, \textit{e.g.}, improved by 1.22$\%$ (Acc-2) and 1.11$\%$ (F1 score) in Table~\ref{tab:table2}, indicating the effectiveness of our proposed approach. 

\begin{table}[t]
\centering
\caption{
Ablation study on the influence of modality on the CMU-MOSEI dataset. The unimodal results are adapted from Mult \cite{tsai2019multimodal}.
}
\resizebox{\columnwidth}{!}{

\begin{tabular}{lcccccc}
\toprule
Input & $MAE\downarrow$ & $Corr\uparrow$ & $Acc_{7}\uparrow$ & $Acc_{2}\uparrow$ & $F_{1}\uparrow$\\
\midrule
T only & 0.653 & 0.631 & 46.5 & 77.4 & 78.2\\
A only & 0.764 & 0.310 & 41.1 & 65.5 & 68.8\\
V only & 0.759 & 0.343 & 43.5 & 66.4 & 69.3\\
TV pairs & 0.544 & 0.765 & 52.45 & 85.36 & 85.32\\
TA pairs & 0.554 & 0.749 & 52.26 & 85.30 & 85.28\\
TV+TA pairs & \textbf{0.532} & \textbf{0.774} & \textbf{53.10} & \textbf{86.27} & \textbf{86.17} \\
\bottomrule
\end{tabular}

}
\label{tab:table6}
\end{table}

\begin{table*}[t]
\centering
\caption{
Ablation study on the influence of the center modality on the CMU-MOSI.
}
\begin{tabular}{lcccccc}
\toprule
Model & $MAE\downarrow$ & $Corr\uparrow$ & $Acc_{7}\uparrow$ & $Acc_{2}\uparrow$ & $F_{1}\uparrow$\\
\midrule
Acoustic-oriented & 0.777 & 0.786 & 42.86 & 84.15 & 84.22 \\
Visual-oriented & 0.745 & 0.789 & 44.61 & 84.30 & 84.31 \\
Text-oriented (Ours) & \textbf{0.714} & \textbf{0.797} & \textbf{46.79} & \textbf{86.28} & \textbf{86.15} \\
\bottomrule
\end{tabular}


\label{tab:table4}
\end{table*}

\begin{table*}[!htp]
\centering
\caption{
Ablation study on the influence of different network modules on the CMU-MOSI.
}

\begin{tabular}{ccccccccc}
\toprule
Text-oriented Module & Gated Module & Joint Learning & $MAE\downarrow$ & $Corr\uparrow$ & $Acc_{7}\uparrow$ & $Acc_{2}\uparrow$ & $F_{1}\uparrow$\\
\midrule
$\checkmark$ & & & 0.746 & 0.776 & 42.17 & 84.30 & 84.24\\
$\checkmark$ & $\checkmark$ & & 0.755 & 0.783 & 42.42 & 84.76 & 84.67\\
$\checkmark$ & $\checkmark$ & $\checkmark$ & \textbf{0.714} & \textbf{0.797} & \textbf{46.79} & \textbf{86.28} & \textbf{86.15} \\
\bottomrule
\end{tabular}


\label{tab:table5}
\end{table*}

\subsection{Ablation Study}
The overall outstanding performance of TCAN has confirmed its superiority. To grasp the significance of various components and ascertain the dominance of the text modality, we conduct systematic ablation experiments on the CMU-MOSI and CMU-MOSEI datasets. The results of these experiments are detailed in the following tables.

We separately remove each modality to investigate the performance of our proposed model. As illustrated in Table~\ref{tab:table6}, we can see that the performance of ``T only'' is remarkably better than ``A only'' and ``V only'', which to some extent implies the importance of text for MSA. When introducing text-oriented cross-attention modules, ``TV pairs'' or ``TA pairs'', the performance has an apparent gain. Then, combining ``TV pairs'' and ``TA pairs'' further improves the recognition performance. In addition, the performance drop is relatively large when the visual modality is removed (comparing ``TV pairs'' and ``TA pairs'' related to ``TV+TA pairs'').

\begin{figure}[t]
    \centering
    \includegraphics[width=\linewidth]{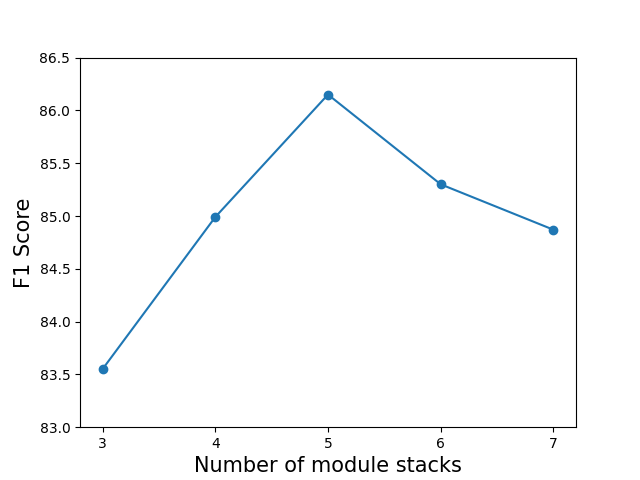}
    \caption{Performance of TCAN with different hyperparameter $N$ on the CMU-MOSI. Here we use $F_{1}$ scores to show the performance.}
    \label{fig:fig4}
\end{figure}

To illustrate the dominance of the text modality, we substitute the other two modalities as the dominant modality for the experiments. As illustrated in Table~\ref{tab:table4}, the acoustic- and visual-oriented models consistently experience noticeable performance degradation. These observations indicate that the text modality possesses richer semantics and less noise, resulting in better feature reinforcement of the other two modalities.

\begin{figure*}[!htp]
    \centering
    \includegraphics[width=\linewidth]{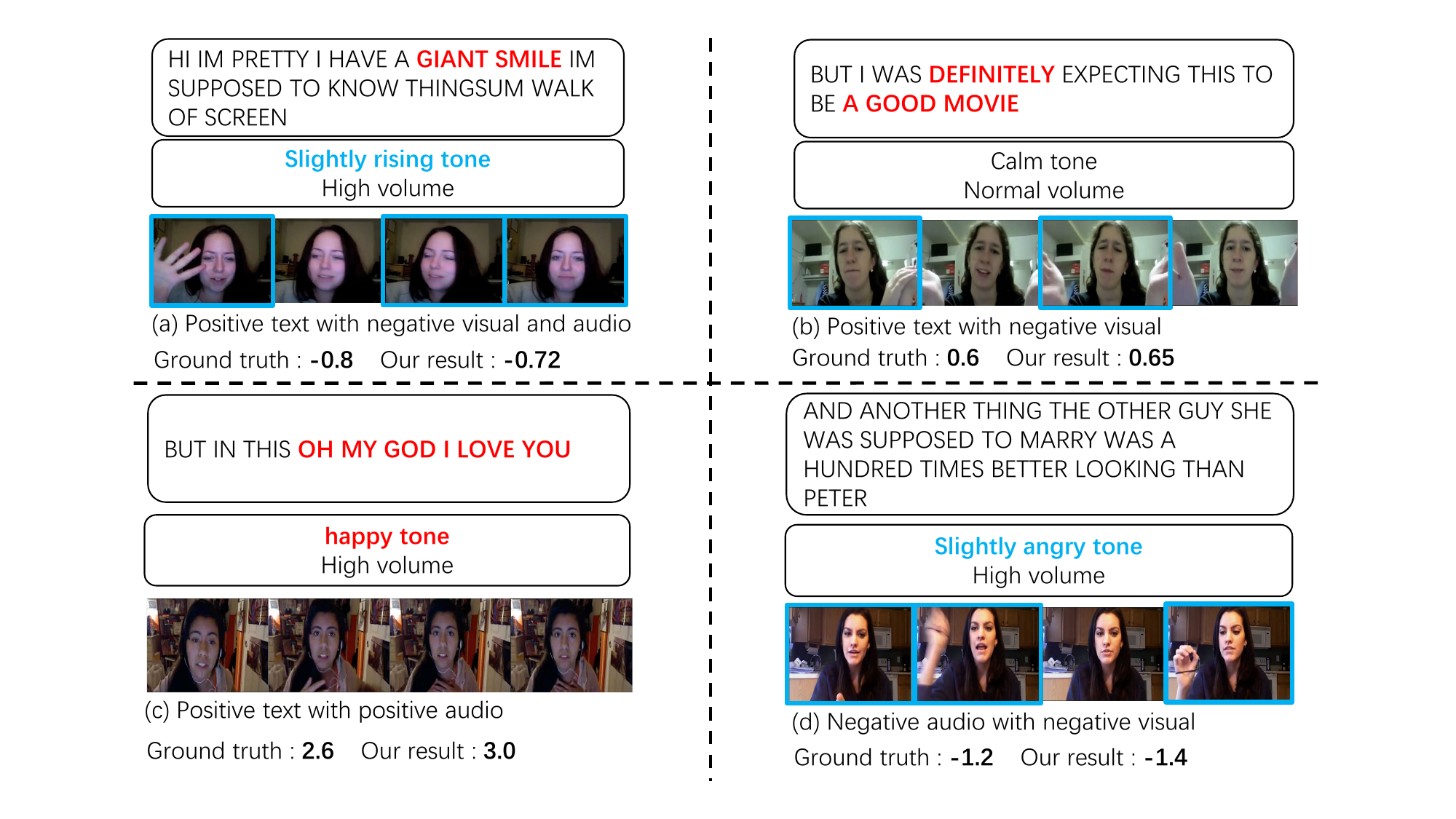}
    \vspace{-10mm}
    \caption{Some representative examples for the visualization analysis of MSA. Red represents positive sentiment and blue represents negative sentiment.
    }
    \label{fig:visualization}
\end{figure*}

We explore the importance of the proposed components by removing the gated mechanism and unimodal joint training separately. We can see that the gain degrades when removing one of the modules. As shown in Table~\ref{tab:table5}, these observations suggest that gated mechanism and joint training modules play an important role in the result.

Finally, to investigate the impact of the number $N$ of the Text-oriented Cross-attention module on the model's performance, we carried out experiments on the CMU-MOSI dataset using varying values for parameter $N$. The results are illustrated in Fig.~\ref{fig:fig4}. As $N$ increases, we observe a pattern in which the F1 scores initially rise and then decline, with the network achieving its optimal performance when $N = 5$. We believed that the larger $N$ could result in better performance. Nevertheless, experiments indicate that an excessive number of layers may restrict the text modality's capacity to guide the other two modalities. We should choose the appropriate network for different datasets, which is exactly what our proposed TCAN can flexibly do. When transitioning our model to a more intricate dataset, we can strategically augment the number of Text-Oriented Cross-modal Attention modules to attain optimal performance. 

\subsection{Visualization Analysis}
To further analyze our proposed framework for MSA, we visualize several misleading examples as shown in Fig.~\ref{fig:visualization}. We use different colors to distinguish sentiments, where red refers to positive sentiment and blue refers to negative sentiment.

In Fig.~\ref{fig:visualization}(a), a positive phase \textit{GIANT SMILE} exists in text modality, but the multimodal label is -0.8, which demonstrates the limitation of only text modality in this case. In fact, there are mostly negative video frames and audio tones that cause the label to be negative. The prediction result of our model is -0.72, which shows that our method can handle the restriction of text modality. In Fig.~\ref{fig:visualization}(b), the visual modality has a slight negative sentiment, while the words \textit{DEFINITELY, A GOOD MOVIE} show that text modality is obviously positive. By leveraging the text modality, TCAN partially compensates for the limitations in sentiment of the acoustic and visual modalities. The result of our model is 0.65, which is similar to the multimodal label. In Fig.~\ref{fig:visualization}(c), there is a positive sentence \textit{OH MY GOD I LOVE YOU} with a happy tone. In Fig.~\ref{fig:visualization}(d), both modalities are slightly negative. In these cases, TCAN can also obtain more accurate results. These cases mentioned above demonstrate the superiority of our model in assigning reasonable weights to different modalities for multimodal sentiment analysis. This also implies that our model will not underestimate the significance of strong modalities while overemphasizing the importance of weak ones.

\section{Conclusion}
\label{sec:conclusion}
In this paper, we propose a cross-attention network designed for text-oriented multimodal sentiment analysis. Our objective is to foster interaction and fusion among three modalities, each comprising unaligned sequences. The proposed methodology is built upon both inter- and intra-modal attention mechanisms. Sequences from the text modality guide the attention of the other two modalities, facilitating the transfer of information across and within different modalities. To address issues related to noise and redundant information, we incorporate a gated mechanism into the model, effectively mitigating these problems. Simultaneously, we jointly train both the multimodal and unimodal tasks to enable the learning of their respective consistencies and differences. Experimental results demonstrate that the proposed approach consistently outperforms the state-of-the-art in multimodal sentiment analysis benchmarks.

\bibliographystyle{IEEEtran}
\bibliography{refs}

\end{document}